# Single-electron charge sensing in self-assembled quantum dots

Haruki Kiyama[1], Alexander Korsch[2], Naomi Nagai[3], Yasushi Kanai[1], Kazuhiko Matsumoto[1], Kazuhiko Hirakawa[3], Akira Oiwa[1,4]

[1] The Institute of Scientific and Industrial Research, Osaka University, 8-1, Mihogaoka, Ibaraki, Osaka 567-0047, Japan
[2] Lehrstuhl für Angewandte Festkörperphysik, Ruhr-Universität Bochum, Universitätsstraße 150, Gebäude NB, D-44780 Bochum, Germany
[3] Institute of Industrial Science, the University of Tokyo, 4-6-1 Komaba, Meguro, Tokyo 153-8505, Japan
[4] Center for Spintronics Research Network, Graduate School of Engineering Science, Osaka University, 1-3 Machikaneyama, Toyonaka, Osaka 560-0043, Japan
E-mail: kiyama@sanken.osaka-u.ac.jp

## ABSTRACT

Measuring single-electron charge is one of the most fundamental quantum technologies. Charge sensing, which is an ingredient for the measurement of single spins or single photons, has been already developed for semiconductor gate-defined quantum dots, leading to intensive studies on the physics and the applications of single-electron charge, single-electron spin and photon−electron quantum interface. However, the technology has not yet been realized for self-assembled quantum dots despite their fascinating quantum transport phenomena and outstanding optical functionalities. In this paper, we report charge sensing experiments in self-assembled quantum dots. We choose two adjacent dots, and fabricate source and drain electrodes on each dot, in which either dot works as a charge sensor for the other target dot. The sensor dot current significantly changes when the number of electrons in the target dot changes by one, demonstrating single-electron charge sensing. We have also demonstrated real-time detection of single-electron tunnelling events. This charge sensing technique will be an important step towards combining efficient electrical readout of single-electron with intriguing quantum transport physics or advanced optical and photonic technologies developed for self-assembled quantum dots.

## Introduction

Self-assembled quantum dots (QDs) have been a fascinating platform for the investigation of microscopic quantum physics and applications to nanoelectronics, spintronics and photonics. In InAs QD-based single-electron transistors [1], a variety of quantum transport experiments have been reported, including electrical control of the spin–orbit interaction [2] and g-factor [3,4], Josephson junction [5,6], spin valve [7,8] and terahertz spectroscopy [9]. In quantum information processing, the coupling between photons and InAs QDs offers key technologies, such as a single-photon source [10], single-spin manipulation [11] and entanglement between spins and photons [12,13]. Moreover, site-selective growth techniques are being developed [14-15], which are indispensable for constructing large-scale quantum devices comprising a number of dots.

For further developments of the potential abilities of self-assembled QDs in quantum information processing, a charge sensing technique for the self-assembled QDs is strongly needed as it has greatly contributed to the development of the state-of-the-art electron spin qubits using gate-defined QDs [16]. Charge sensing has been realized by using a quantum point contact [17] or a single electron transistor [18] fabricated near the dots as a sensor. Because of the capacitive coupling between the dot and the sensor, the sensor conductance changes with the number of electrons in the dots. This technique works even if dot conductance is too small to measure, enabling detection of single-electron tunnelling events in real-time [18-20]. In addition, this real-time charge sensing allows for measurement of other physical quantities by converting them into electron charge: single-photon detection has been demonstrated by detecting single photo-excited electrons [21,22], and the readout of single-electron spin has been implemented by detecting spin-dependent tunnelling events [23-26]. Moreover, a few works on charge sensing have been reported for vertical QDs [27], carbon nanotube QDs [28] and nanowire QDs [29-31] by placing a sensor near a QD or by connecting a sensor and a QD with a floating gate. However, this has not yet been achieved for self-assembled QDs. Realizing charge sensing in a single-electron transistor based on a self-assembled QD may be challenging because the metal electrodes directly contacting the QD may effectively screen the single-electron charge in the QD [6].

In this work, we report single-electron charge sensing experiments in InAs self-assembled QDs by using another adjacent QD as a sensor. Metal electrodes contacting the QDs are made narrow for reducing the screening effect. The capacitive coupling between the two QDs is large enough to show the distinct change in the dot current induced by the single-electron charging in the adjacent dot. We also demonstrate charge sensing in real-time at the dot-reservoir resonance having the tunnel rate lower than the measurement bandwidth.

## Results
**Device.**
Two samples, A and B, studied in this work consist of two uncapped self-assembled InAs QDs, each contacted by a pair of Ti/Au (10/20 nm) electrodes as source and drain [see Fig. 1(a)]. The two QDs have a diameter and height of approximately 100 nm and 20 nm, respectively. These dots are separated by approximately 150 nm from centre-to-centre. The source-drain electrode has a nanogap separation of approximately 50 nm and a width of approximately 50 nm. The latter is intentionally made narrow compared to the devices used in preceding studies [1-9, 32], in order to reduce the screening effect [6] and hence enhance the charge sensitivity of the dots. Ti/Au electrodes surrounding the dots and a degenerately Si-doped GaAs layer buried 300 nm below the surface are used as local side gates and a global back gate, respectively, to control the electrostatic potential of the dots. For sample A, the 20 μm × 20 μm region around the dots is covered by a 50-nm-thick SiN$_x$ film to increase the capacitance between the two dots. In the following, we denote the two QDs as QD$_i$ ($i$ = 1,2), and the side gates as SG$_{iL}$ and SG$_{iR}$ for each sample, as illustrated in Fig. 1(a). We measure the current through QD$_i$, $I_i$, at the source–drain bias voltage, $V_{SDi}$, across the dot, the side-gate voltages, $V_{SGiL}$ and $V_{SGiR}$, and the back-gate voltage, $V_{BG}$. All measurements discussed below have been performed in a dilution refrigerator at a base temperature of 20 mK and an electron temperature of 290 mK.

**Basic characterization of devices.**
Figure 1(b) shows the differential conductance of QD$_2$ in sample A, $dI_2/dV_{SD2}$, as a function of $V_{SD2}$, and the side-gate voltage $V_{SG2} = V_{SG2L} = V_{SG2R}$ at $V_{BG}$ = 0 V. We observe a series of diamond-shaped Coulomb blockade regions. Some of the diamonds are truncated because of inelastic co-tunnelling processes. From the aspect ratio of the diamonds, we evaluate the lever-arm factor, $\alpha$, which converts the gate voltage to the change in the electrochemical potential of a QD. We find that $\alpha$ varies with the electron number in the dot, as implied by different aspect ratios of the diamonds in Fig. 1(b). We suppose that $\alpha$ depends on the spatial distribution of the wave function in the dot and thus the screening effect, which may change with the electron number. In Fig. 1(c), we show averaged $\alpha$ values of each dot in samples A and B for $V_{SG1} = V_{SG1L} = V_{SG1R}$, $V_{SG2}$ and $V_{BG}$. Error bars indicate the range of the $\alpha$ values evaluated for single diamonds. For the side gates, the values of $\alpha$ are larger than 40 meV/V, which is an order of magnitude larger than those reported in preceding works [2,3]. These large $\alpha$ values of the side gates are attributed to the gates' locations close to the dots, increased number of side gates, and the reduced screening effect by the narrow source and drain electrodes. The $\alpha$ values of the side gates in samples A are larger than those in sample B, whereas the $\alpha$ values of the back gates are almost the same for both samples. This may indicate the increase in the capacitance between the dot and the side gates by the SiN$_x$ dielectric layer; this is not, however, conclusive because of the electron number dependence of $\alpha$.

**Charge sensing experiments.**
To investigate the transport properties of one dot in response to single-electron charging in the other, we measure $I_1$ and $I_2$ simultaneously. Figures 2(a) and 2(b) show $I_1$ and $I_2$, respectively, in sample A as a function of $V_{SG1R}$ and $V_{SG2L}$ at $V_{BG}$ = 0 V and $V_{SD1} = V_{SD2}$ = 70 μV. We observe two Coulomb peak ridges in each dot in the measured side-gate voltage range. These ridges exhibit a finite slope because each dot has capacitive couplings to both the side gates. Each ridge shows a distinct shift at the gate voltage where the two ridges in the different dots intersect each other. The gate voltage of the ridges shifts towards positive when the electron number in the other dot increases, resulting in a honeycomb pattern [see also Fig. 2(c)] typical of double QDs with an inter-dot capacitive coupling [33]. These features indicate that these shifts of ridges result from the capacitive coupling between the two dots, and thus, the charge sensing of either dot by using the other dot as a sensor is achieved. Note that the inter-dot tunnel coupling is negligibly small. We observed similar charge sensing features in sample B as shown in Figs. 2(d) and 2(e).

We evaluate the change in the electrochemical potential of the dot, $\Delta\mu$, induced by single-electron charging in the other dot at the gate voltage conditions $P_j$ ($j = 1$–$4$) shown in Fig. 2(a) for sample A and $Q_k$ ($k = 1$–$3$) shown in Fig. 2(d) for sample B. For sample A, $\Delta\mu$ is $224 \pm 24$ µeV at $P_1$ and $P_2$, and $110 \pm 24$ µeV at $P_3$ and $P_4$. For sample B, $\Delta\mu$ is $114 \pm 23$ µeV at $Q_1$ and $154 \pm 23$ µeV at $Q_2$ and $Q_3$. For each of $P_j$ and $Q_k$, the values of $\Delta\mu$ evaluated from $I_1$ [Figs. 2(a) and 2(d)] are almost the same as those evaluated from $I_2$ [Figs. 2(b) and 2(e)]. It is difficult to compare $\Delta\mu$ directly between samples A and B because $\Delta\mu$ reflects the details in the wave function geometry in both dots as discussed above for the lever-arm factors in Fig. 1(c). It has been reported that asymmetrically applied side-gate voltage affects the lateral position and the extension of the wave function because of the modulation of the lateral confinement potential of the QDs [4,5]. As seen in Fig. 2(b), for $QD_2$ in sample A, the electronic state responsible for the lower Coulomb peak ridge couples more strongly to $SG_{2L}$ than the state for the upper ridge. This implies that the orbital wave function in $QD_2$ is located closer to $SG_{2L}$ and thus farther from $QD_1$, which is consistent with the smaller $\Delta\mu$ for the lower ridge than the upper one.

**Demonstration of real-time charge sensing.**

We demonstrate the real-time detection of single-electron tunnelling events. Figure 3(a) shows $I_1$ in sample B as a function of $V_{SG1} = V_{SG1L} = V_{SG1R}$ and $V_{SG2L}$ for different electron numbers in both $QD_1$ and $QD_2$ from Figs. 2(d) and 2(e) with $V_{BG} = -0.5$ V, $V_{SD1} = 200$ µV and $V_{SD2} = 0$ µV. We observe the shift in the Coulomb peak ridge for $I_1$ similar to that observed in Figs. 2(a) to 2(e), indicating the change in the electron number of $QD_2$ between $N_2$ and $N_2+1$. The amplitude of $I_2$ is smaller than the noise floor of approximately 20 fA in the same gate-voltage range, suggesting small tunnel coupling between $QD_2$ and the leads. We measure $I_1$ in real-time at the resonance of the charge state transition in $QD_2$ in Fig. 3(a). For the charge sensing measurements, the gate voltages are set such that the sensor $QD_1$ is always set on one side of a Coulomb peak. The measurement bandwidth and the sampling frequency are 5 kHz and 2 kHz, respectively. Figure 3(b) shows the real-time traces of $I_1$ measured at slightly different gate voltage conditions for $QD_2$ along the red line in Fig. 3(a). We observe distinct random telegraph signals between $I_1$ values of ~ 0.2 nA and ~ 0.5 nA. These low and high $I_1$ signal levels correspond to the $N_2$ and $N_2+1$ charge states in $QD_2$, respectively, indicating the detection of single-electron tunnelling events in and out of $QD_2$. The charge-sensing signal amplitude is ~ 0.32 nA and the noise amplitude is ~ 0.08 nA, resulting in the single-to-noise ratio of ~ 4. To further confirm the real-time charge sensing, we analyse the gate voltage dependence of the fraction of the $N_2$ and $N_2+1$ charge states. When the electrochemical potential of $QD_2$, $\mu_{QD2}$, changes with the gate voltages, the fraction of the $N_2$ and $N_2+1$ charge states also changes, reflecting the energy difference between $\mu_{QD2}$ and the thermally broadened energy distributions of electrons in the source and drain electrodes. In Fig. 3(b), as $V_{SG2L}$ decreases across the charge state resonance in $QD_2$ from top to bottom $I_1$ traces, the fraction of the $N_2+1$ state decreases as expected for increasing $\mu_{QD2}$. Figure 3(c) shows the $N_2+1$ state fraction as a function of $\mu_{QD2}$. The estimation of the lever-arm factor in this condition is described in Supplementary Information. Numerical fitting using the Fermi distribution function gives an electron temperature of 280 ± 20 mK, which is in good agreement with that estimated from the Coulomb peak width. We define $t_{in}$ and $t_{out}$ for the dot as the lengths of time it resides at the $N_2+1$ and $N_2$ charge states, respectively, and show their histograms in Figs. 3(c) and 3(d). Each histogram shows a single exponential distribution. By fitting them to $\exp(-\Gamma_{in(out)}t_{in(out)})$, where $\Gamma_{in(out)}$ is the rate of electron tunnelling into (out of) the dot, we obtain the tunnel coupling $\gamma = \Gamma_{in}+\Gamma_{out} = 84 \pm 2$ Hz between the dot and the lead. The utility of the real-time charge sensing will be further improved by the ability to independently control the tunnel rate and the charge state in individual QDs. This will be viable by changing voltages on both side and back gates simultaneously so as to modulate the overlap between the lead state and the electronic wave function in the QD while keeping the electrochemical potentials of the QD unchanged [5]. The control of the tunnel rate for arbitrary charge states is to be investigated in our devices in future experiments.

## Summary and Prospect

In summary, we have demonstrated charge sensing experiments in InAs self-assembled quantum dots by using one of two adjacent dots as a target and the other as a sensor. We have observed distinct shifts in the Coulomb peak ridges in the sensor dot when the electron number changes by one in the target dot, which is a signature of single-electron charge sensing. We have also demonstrated real-time detection of single-electron tunnelling events, which is an ingredient for the measurement of single spins or single photons. The charge sensing technique presented in this work will be applicable for self-assembled QDs made of other materials, such as GaN[34] and SiGe[35]; this would bring opportunities to investigate intriguing physics of charge and spin in self-assembled QDs. Moreover, the technology can be applicable for self-assembled multiple QD systems [16, 36] and QDs coupled to superconductors [5,6] or ferromagnets [7,8]. Our demonstration of the charge sensing in self-assembled QDs will be an important step towards combining efficient electrical readout of single electron with a variety of transport phenomena and advanced optical and photonic technologies.

## Methods

**Device fabrication.**

Uncapped InAs self-assembled QDs are grown by molecular beam epitaxy in the Stranski–Krastanov mode on a semi-insulating (001) GaAs substrate. The growth layers consist of a 300-nm-thick degenerately Si-doped GaAs layer, used as the back gate, followed by a 100-nm-thick $Al_{0.3}Ga_{0.7}As$ barrier layer and a 200-nm-thick undoped GaAs buffer layer. Among the randomly positioned QDs with various sizes, we identify using scanning electron microscope two QDs with size and position suitable for subsequent fabrication. In this work, we choose QDs having diameter and height of approximately 100 nm and 20 nm, respectively, and separated by approximately 150 nm from centre-to-centre. A pair of source and drain electrodes and side-gate electrodes are fabricated on each dot using electron beam lithography and electron beam evaporation of Ti/Au (10 nm/20 nm) [Fig. 1(a)]. Prior to the evaporation, an oxidized layer on the dot surface is removed using in situ Ar plasma etching. For sample A, a 50-nm-thick $SiN_x$ film is deposited by the catalytic chemical vapour deposition method.

**Measurement details.**

Measurements are performed in a dilution refrigerator with a base temperature of 20 mK. The measurement lines are filtered with a two-stage RC low-pass filter with a cut-off frequency of 1MHz. The electron temperature is estimated to be approximately 290 mK from the width of the narrowest Coulomb peaks in our samples.

**Acknowledgements**

We thank S. Baba for help with device fabrication, and T. Hirayama, R. Shikishima for useful discussions. This work was supported by Grant-in-Aid for Young Scientists B (No. 15K17681), Grants-in-Aid for Scientific Research S (No. 17H06120, No. 26220710), Scientific Research A (16H02333), Innovative Areas "Nano Spin Conversion Science" (No. 26103004), Innovative Areas "Science of hybrid quantum systems" (No. 15H05868), Project for Developing Innovation Systems of MEXT, CREST, Japan Science and Technology Agency (JST) (JPMJCR15N2), the Murata Science Foundation, the Asahi Glass foundation, Dynamic Alliance for Open Innovation Bridging Human, Environment and Materials from MEXT, "Nanotechnology Platform Project (Nanotechnology Open Facilities in Osaka University)" of Ministry of Education, Culture, Sports,

**Author Contributions**

H.K. fabricated the devices, performed the measurements, analysed the results and wrote the manuscript. A.K. contributed to the device fabrication and the measurements. N.N. and K.H. grew the uncapped InAs QDs. Y.K. and K.M deposited the $SiN_x$ layer. A.O. contributed to the interpretation of results and supervised the work. All authors contributed to the discussions and writing of the manuscript.

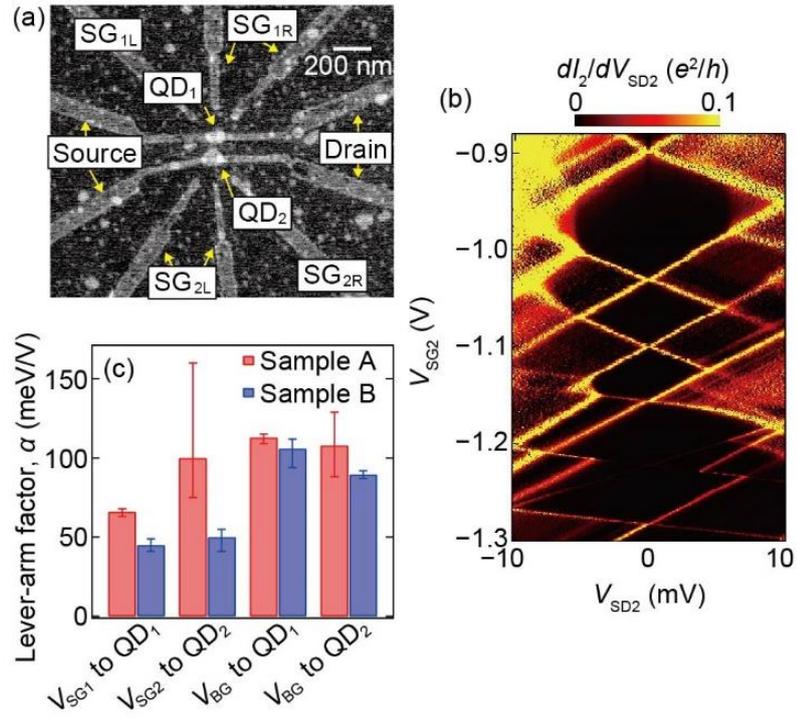

Figure 1. (a) Scanning electron micrograph of a device similar to the measured device. (b) The differential conductance of QD$_2$, $dI_2/dV_{SD2}$, in sample A as a function of $V_{SD2}$ and $V_{SG2}$ at $V_{BG}$ = 0 V. (c) Lever-arm factor $\alpha$ of the side gates and the back gates for samples A and B.

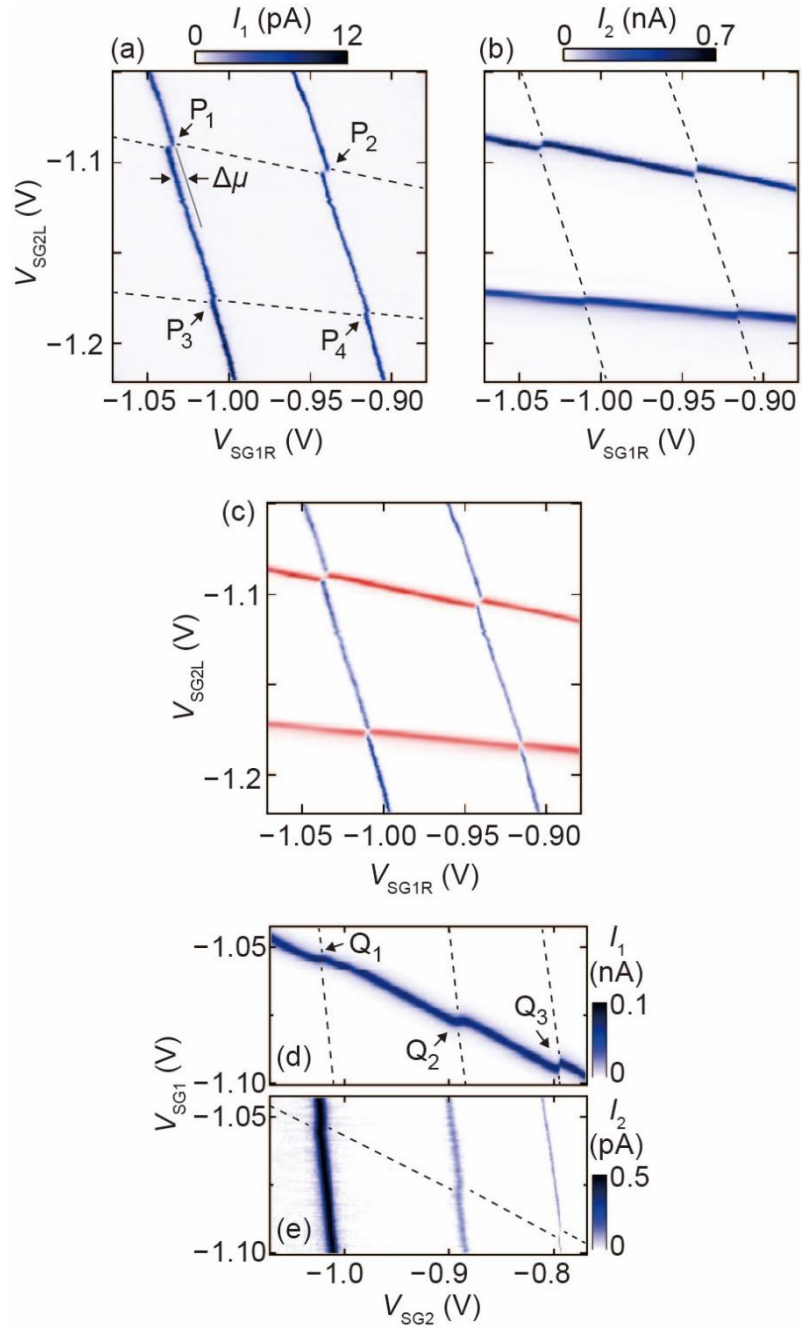

Figure 2. (a), (b) Intensity plots of $I_1$ (a) and $I_2$ (b) in sample A as a function of $V_{SG1R}$ and $V_{SG2L}$ at $V_{BG} = 0$ V and $V_{SD1} = V_{SD2} = 70$ μV. The dashed lines in (a) and (b) represent the positions of the Coulomb peak ridges in $QD_2$ and $QD_1$, respectively. Capacitive coupling between the two QDs and thus charge sensing features are observed at the gate voltage conditions denoted as $P_j$ ($j$ = 1–4), where the two ridges in different QDs intersect. (c) Superposition plots of (a) and (b) showing the honeycomb pattern. (d), (e) Intensity plots of $I_1$ (d) and $I_2$ (e) in sample B as a function of $V_{SG1}$ and $V_{SG2}$ at $V_{BG} = 0$ V and $V_{SD1} = V_{SD2} = 70$ μV, showing the charge sensing features at $Q_k$ ($k$ = 1–3).

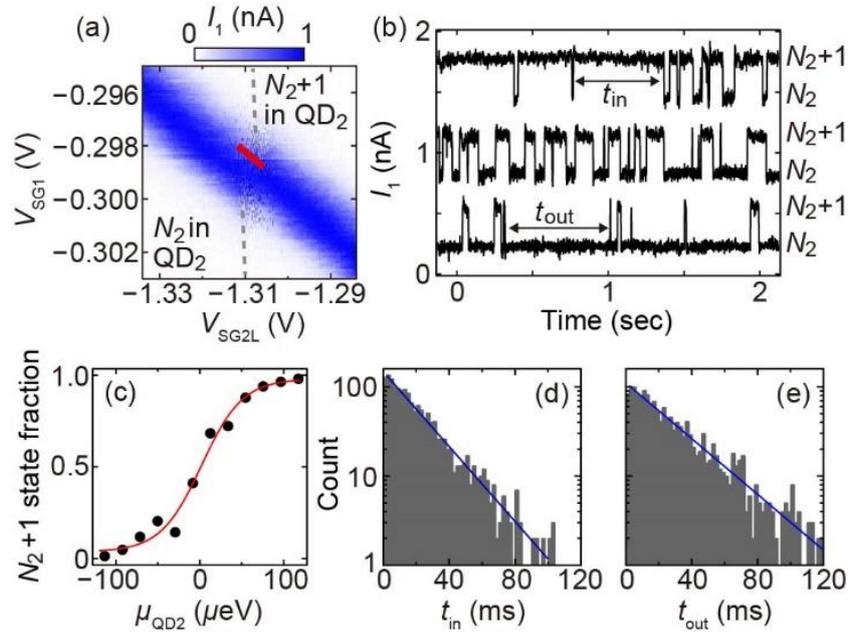

Figure 3. (a) Intensity plot of $I_1$ in sample B as a function of $V_{SG1} = V_{SG1L} = V_{SG1R}$ and $V_{SG2L}$ at $V_{BG} = -0.5$ V, $V_{SD1} = 200$ μV and $V_{SD2} = 0$ μV. (b) Real-time traces of $I_1$ measured at $V_{SG2L} = -1.306$ V (top), $-1.309$ V (middle), and $-1.312$ V (bottom) along the red line in (a). Each curve is offset by 0.6 nA for clarity. (c) $N_2+1$ state fraction as a function of $\mu_{QD2}$. A red curve is a fit to the data with the Fermi distribution function. (d), (e) Histograms of $t_{in}$ (d) and $t_{out}$ (e) obtained at $V_{SG2L} = -1.306$ V, the same condition as the middle $I_2$ trace in (b).

# Single-electron charge sensing in self-assembled quantum dots

Haruki Kiyama[1], Alexander Korsch[2], Naomi Nagai[3], Yasushi Kanai[1], Kazuhiko Matsumoto[1], Kazuhiko Hirakawa[3], Akira Oiwa[1,4]

[1] *The Institute of Scientific and Industrial Research, Osaka University, 8-1, Mihogaoka, Ibaraki, Osaka 567-0047, Japan*

[2] *Lehrstuhl für Angewandte Festkörperphysik, Ruhr-Universität Bochum, Universitätsstraße 150, Gebäude NB, D-44780 Bochum, Germany*

[3] *Institute of Industrial Science, the University of Tokyo, 4-6-1 Komaba, Meguro, Tokyo 153-8505, Japan*

[4] *Center for Spintronics Research Network, Graduate School of Engineering Science, Osaka University, 1-3 Machikaneyama, Toyonaka, Osaka 560-0043, Japan*

E-mail: kiyama@sanken.osaka-u.ac.jp

**Estimation of the lever-arm factor for a quantum dot weakly tunnel-coupled to reservoirs**

In Fig. 3(c) in the main text, we show the $N+1$ state fraction of the quantum dot $QD_2$ in sample A as a function of the dot chemical potential $\mu_{QD2}$. Although the current through $QD_2$, $I_2$, is too small to measure at the real-time charge sensing condition, we estimate the lever-arm factor of the side-gate voltage $V_{SG2L}$ to $\mu_{QD2}$ from the charge sensing measurement of $QD_2$. Figures S1(a) to (c) show the current through $QD_1$, $I_1$, as a function of side-gate voltages $V_{SG1}$ and $V_{SG2L}$ at different source and drain bias voltages across $QD_2$, $V_{SD2}$. The gate voltage conditions are slightly changed from those in Fig. 3(a) supposedly because of the charge redistribution around the QDs. The Coulomb peak resonance in $QD_2$ appears at the shift of the $I_1$ Coulomb peak as indicated by the red arrows. As $V_{SD2}$ decreases, the $V_{SG2L}$ value of the $QD_2$ resonance decreases without significant broadening of the resonance. This indicates that the source electrode has larger tunnel coupling with $QD_2$ than the drain electrode, and that electron tunnelling occurs dominantly between $QD_2$ and the source electrode. Note that $V_{SG1}$ is slightly increased with decreasing $V_{SD2}$ to tune the chemical potential of $QD_1$, which hardly changes $\mu_{QD2}$. Figure S2 shows the $V_{SG2L}$ value for the $QD_2$ resonance as a function of $V_{SD2}$. From a linear fit to this data, we estimate the lever-arm factor of $V_{SG2L}$ to $\mu_{QD2}$ to be $21 \pm 1$ meV/V by assuming that the source and the drain electrode have almost the same capacitance with $QD_2$.

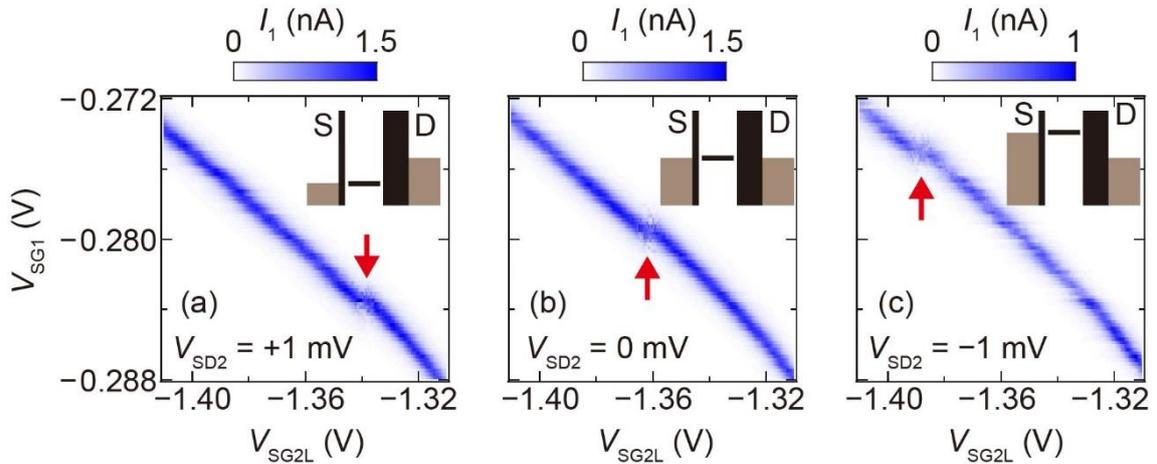

Figure S1. Intensity plot of $I_1$ in sample B as a function of $V_{SG1} = V_{SG1L} = V_{SG1R}$ and $V_{SG2L}$ at (a) $V_{SD2} = +1$ mV, (b) $V_{SD2} = 0$ mV, and (c) $V_{SD2} = -1$ mV. The back-gate voltage is $V_{BG} = -0.5$ V, and the source and drain bias voltage across $QD_1$ is $V_{SD1} = 70$ μV. Insets schematically illustrate the energy diagram at the Coulomb peak resonance in $QD_2$ for each $V_{SD2}$.

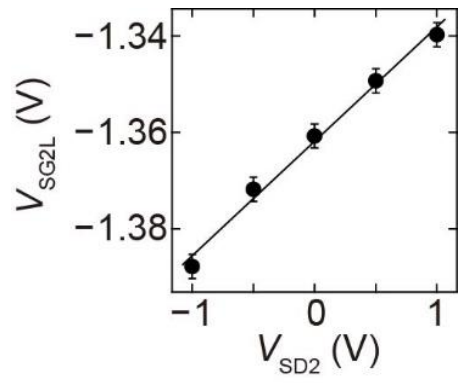

Figure S2. $V_{SG2L}$ values for the QD$_2$ resonance as a function of $V_{SD2}$. The solid line represents a linear fit to the data.